\documentclass[final,5p,times,twocolumn,numbers,sort&compress]{elsarticle}

\usepackage{booktabs}
\usepackage{amsmath}
\usepackage{caption}
\usepackage{amssymb}
\usepackage{lipsum}
\usepackage{amsmath}
\usepackage{multirow}
\usepackage{amssymb}
\usepackage{hyperref}
\usepackage{xcolor}
\usepackage{bm}
\usepackage{amsthm}
\usepackage[normalem]{ulem}

\renewcommand{\sout}{\bgroup \color{red} \ULdepth=-.5ex \ULset}

\journal{Physics Letters B}

\begin{document}

\begin{frontmatter}

\title{Effects of mean-field momentum dependence on
pion production in intermediate-energy heavy-ion collisions}

\author[label1,label2]{Xin Li\corref{cor1}}
\ead{lixin20223@stu.htu.edu.cn}
\affiliation[label1]{organization={College of Physics},
             addressline={Henan Normal University},
             city={Xinxiang},
             postcode={453007},
             country={China}}
\author[label2]{Si-Pei Wang\corref{cor1}}
\ead{sjtuwsp@gmail.com}
\affiliation[label2]{organization={State Key Laboratory of Dark Matter Physics}, organization={Key Laboratory for Particle Astrophysics and Cosmology (MOE)}, organization={and Shanghai Key Laboratory for Particle Physics and Cosmology}, organization={School of Physics and Astronomy}, organization={Shanghai Jiao Tong University}, city={Shanghai}, postcode={200240}, country={China}}
\author[label5,label4]{Zhen Zhang\corref{cor1}}
\ead{zhangzh275@mail.sysu.edu.cn}
\affiliation[label5]{organization={Sino-French Institute of Nuclear Engineering and Technology},
	addressline={Sun Yat-sen University},
	city={Zhuhai},
	postcode={519082},
	country={China}}
\author[label6,label4]{Rui Wang\corref{cor1}}
\ead{rui.wang@lns.infn.it}
\affiliation[label6]{organization={Istituto Nazionale di Fisica Nucleare (INFN) },
	addressline={ Laboratori Nazionali del Sud},
	city={Catania},
	postcode={I-95123},
	country={Italy}}
\author[label1,label3,label4]{Jie Pu\corref{cor1}}
\ead{wlpujie@126.com}
\affiliation[label3]{organization={Institute of Nuclear Science and Technology},
	addressline={Henan Academy of Sciences},
	city={Zhengzhou},
	postcode={450046},
	state={},
	country={China}}
\affiliation[label4]{organization={Shanghai Research Center for Theoretical Nuclear Physics},
	city={NSFC and Fudan University},
	postcode={Shanghai},
	state={200438},
	country={China}}
\author[label1,label3,label4]{Chun-Wang Ma\corref{cor1}}
\ead{machunwang@126.com,machw@htu.edu.cn}
\author[label2]{Lie-Wen Chen\corref{cor1}}
\cortext[cor1]{Corresponding author.}
\ead{lwchen@sjtu.edu.cn}

\begin{abstract}
Pion production in heavy-ion collisions at intermediate energies provides an important probe of the collision dynamics and nuclear matter equation of state, especially the high-density behavior of the symmetry energy.
Using the lattice Boltzmann-Uehling-Uhlenbeck transport model
with a recently developed nuclear effective interaction based on the so-called N$5$LO Skyrme pseudopotential, we investigate the effects of the momentum dependence of nucleon mean-field potentials on the pion production in Au+Au collisions at a beam energy of $1.23$~GeV/nucleon.
We find that a stronger momentum dependence, for which the nucleon mean-field potentials increase faster with momentum, generally suppresses pion production. This feature can be understood in terms of the mean-field-induced modification of nucleon high-momentum phase space during the compression stage: a stronger momentum dependence can reduce the relative fraction of high-momentum nucleons in heavy-ion collisions, thereby suppressing the production of $\Delta$ resonances and pions.

\end{abstract}




\end{frontmatter}




\section{Introduction}
\label{introduction}
Heavy-ion collisions (HICs) at intermediate energies (from $\sim 100$ MeV/nucleon to $\sim 1$ GeV/nucleon) provide a unique tool to explore the equation of state~(EOS) of dense nuclear matter, which lies at the core of both nuclear physics and astrophysics~\cite{Li:1997px,Lattimer:2000kb,Danielewicz:2002pu,Lattimer:2004pg,Steiner:2004fi,Baran:2004ih,Li:2008gp,Oertel:2016bki,Sorensen:2023zkk,Chen:2024aom}.
As the lightest hadrons, pions are among the most abundant reaction products in intermediate-energy HICs and carry valuable information on the collision dynamics and nuclear matter EOS.
In particular, the charged-pion ratio was proposed as a sensitive probe of the high-density behavior of the nuclear symmetry energy~\cite{Li:2002qx}, and in the past decades, extensive efforts have been devoted to constraining the density dependence of the symmetry energy from transport model analyses of pion observables~\cite{Xiao:2008vm,Feng:2009am,Xie:2013np,Xu:2013aza,Hong:2013yva,Song:2015hua,Zhang:2017mps,Zhang:2018ool,Ikeno:2016xpr,SpiRIT:2021gtq}.
However, pronounced model dependence, leading in some cases to controversial conclusions, has been found in various transport model analyses of charged-pion production in Au+Au collisions measured by the FOPI Collaboration~\cite{FOPI:2006ifg}.
These ambiguities motivate further investigations of pion-production mechanisms in HICs, including $\Delta$ potentials~\cite{Li:2015hfa,Cozma:2021tfu,Ikeno:2023cyc}, pion potentials~\cite{Hong:2013yva, Guo:2014fba, Feng:2015vra, Cozma:2016qej,Zhang:2017mps}, threshold effects~\cite{Ferini:2005del,Song:2015hua}, energy-conservation effects~\cite{Cozma:2014yna,Zhang:2017nck,Ikeno:2023cyc}, and short-range correlations~\cite{Li:2014vua, Yong:2015gma}.
To further quantify uncertainties and improve the predictive power of transport models at these energies, the Transport Model Evaluation Project (TMEP) has also been initiated~\cite{TMEP:2016tup,TMEP:2017mex,TMEP:2021ljz,TMEP:2022xjg,TMEP:2019yci,TMEP:2023ifw}.

Recently, the HADES Collaboration has measured charged-pion production in Au+Au collisions at $\sqrt{s_{\rm NN}}=2.4~\mathrm{GeV}$ (corresponding to beam energy $E_{\rm{beam}}$ = $1.23$ GeV/nucleon) with high-statistics data~\cite{HADES:2020ver}.
These measurements provide an important benchmark for transport model comparisons and can help refine theoretical descriptions of pion production in HICs.
Nevertheless, substantial differences between transport-model predictions and the HADES data on pion yields have been observed~\cite{HADES:2020ver}.
Subsequent studies~\cite{Godbey:2021tbt,Kummer:2023hvl} have suggested that in-medium suppression of the $\Delta$-production cross section can help reconcile transport-model calculations with the HADES results.
Given the complexity of pion-production mechanisms, further investigation into pion production in HICs at HADES energies remains crucial, especially for constraining nuclear matter EOS with the new data.

Nucleon mean-field potentials are among the most important ingredients in transport models. A key feature of these potentials is their momentum dependence, which arises from the intrinsic momentum dependence of nuclear force, finite-range exchange terms, and other possibile contributions~\cite{Decharge:1979fa,Wiringa:1988jt}.
This momentum dependence, which has been clearly demonstrated by nucleon optical-potential measurements, governs essential phenomena ranging from single-particle motion to collective dynamics in nuclear matter~\cite{Danielewicz:2002pu,Liu:2025pzr,Wang:2024xzq,Danielewicz:1999zn,Li:2018lpy,Rizzo:2003if}, and therefore plays a crucial role in the non-equilibrium evolution of nuclear systems.
In HICs, a stronger momentum dependence (i.e., nucleon potential rises more steeply with momentum) is expected to lower the fraction of high-momentum nucleons during compression through energy-conservation effects, in a manner analogous to how the symmetry energy regulates the isospin composition of dense matter in \textit{isospin fractionation}~\cite{Li:2000bj,Xu:1999bs,Li:2002yda}.
This modification of the nucleon momentum distribution, driven by the momentum dependence of nucleon mean field potentials, can significantly influence reaction dynamics and particle production.

In this work we investigate how the momentum dependence of nucleon mean-field potentials influences pion production in Au+Au collisions at HADES energies, by using the lattice Boltzmann-Uehling-Uhlenbeck (LBUU) transport model~\cite{Wang:2018yce,Wang:2019ghr,Wang:2020ixf} incorporated with the extended Skyrme interactions named N$5$LO Skyrme pseudopotential~\cite{Wang:2024xzq}. Our results indicate that the  momentum dependence of nucleon mean-field potentials can significantly influence pion production.

This paper is organized as follow. A brief introduction to the model, particularly the incorporated extended Skyrme interactions with different momentum dependence and the modeling of pion production, is provided in Section~\ref{Sec2}. We present and discuss our results in Section~\ref{Sec3}, and conclude with a summary and outlook in Section~\ref{Sec4}.

\section{Methodology}\label{Sec2}
In this work, we employ the LBUU transport model~\cite{Wang:2018yce,Wang:2019ghr,Wang:2020ixf} to study the nucleon momentum fractionation effects on pion production in Au+Au collisions at $\sqrt{s_{\rm{NN}}}=2.4$ GeV~\cite{HADES:2020ver}.
Within the LBUU model, the time evolution of the one-body phase-space distribution function (Wigner function) $f_{\tau} = f_{\tau}(\vec{r}, \vec{p}, t)$ obeys the Boltzmann-Uehling-Uhlenbeck~(BUU) equation:
\begin{equation}
	\left( \partial_t + \nabla_p \epsilon_{\tau} \cdot \nabla_r - \nabla_r \epsilon_{\tau} \cdot \nabla_p \right) f_{\tau} = I^{\text{coll}}_{\tau}[f_n, f_p, f_{\Delta}, f_{\pi}].
\end{equation}
Here, $\tau$ represents particle species, including neutrons ($n$), protons ($p$), the $\Delta(1232)$ resonances, higher-lying resonances $N^*$ and $\Delta^*$ up to $N(1720)$ and $\Delta(1950)$, and pions ($\pi$). The  single-particle energy is denoted by $\epsilon_{\tau}$, and the collision integral $I^{\mathrm{coll}}_{\tau}$ includes the following processes and their inverse reactions:
\begin{eqnarray*}
	&&N+N\rightarrow N+N, ~~N+N\rightarrow N+R  \\
	&&N^*(\Delta^*) \rightarrow \pi+ \Delta,~~ R \rightarrow \pi+ N,
\end{eqnarray*}
where $R$ denotes resonances including $\Delta$, $N^*$ and $\Delta^*$.
In the LBUU model, the mean-field evolution is treated with the lattice Hamiltonian method~\cite{Lenk:1989zz,Wang:2019ghr}, while the collision integral is handled via the stochastic collision approach~\cite{Danielewicz:1991dh, Xu:2004mz}. For the initial condition, the Thomas-Fermi initialization~\cite{Danielewicz:1999zn,Wang:2019ghr,Gaitanos:2010fd} provides the nuclear ground state as a static solution of the BUU equation, thereby ensuring evolutionary stability. The present framework of solving the BUU equation has been successfully applied to investigate nuclear collective motions in heavy nuclei~\cite{Wang:2019ghr,Wang:2020xgk,Song:2021hyw,Song:2023fnc}, light-nuclei yields~\cite{Wang:2023gta,Wang:2025wim} and proton collective flows~\cite{Wang:2024xzq} in HICs at intermediate energies. A more detailed descriptions of the present framework can be found in Ref.~\cite{Wang:2020ixf}.

\begin{table}
	\caption{The macroscopic characteristic quantities of \text{SP10}, \text{SP10-FL} and \text{SP10-MID}. All interactions share the common parameters of $\rho_{0}$ = $0.160$ $\text{fm}^{-3}$, $E_{0}(\rho_0)$ = $-16.0$ \text{MeV}, $K_{0}$ = $230$ \text{MeV}, $J_{0}$ = $-383$ \text{MeV}, $I_{0}$ = $1818.9$ \text{MeV}, $H_{0}$ = $-12065$ \text{MeV}, $E_{\rm{sym}}(\rho_0)$ = $30$ \text{MeV}, $L$ = $45$ \text{MeV}, $K_{\rm{sym}}$ = $-110$ \text{MeV}, $J_{\rm{sym}}$ = $700$ \text{MeV}, $I_{\rm{sym}}$ = $-2458.5$ \text{MeV}, $H_{\rm{sym}}$ = $19663$ \text{MeV}.}
	\label{tab:sp_properties}
	\small 
	\begin{tabular}{cccc}
		\toprule
		& SP10 & SP10-FL & SP10-MID  \\
		\midrule
		$a_0 \, (\text{MeV})$ & $-64.97$ & $-64.44$ & $-52.84$ \\
		$a_2 \, (\text{MeV fm}^2)$ & 7.104 & 6.829 & 0 \\
		$a_4 \, (\text{MeV fm}^4)$ & $-0.1628$ & $-0.1719$ & 0 \\
		$a_6 \, (\text{MeV fm}^6)$ & $1.731\ \times 10^{-3}$ & $1.908 \times 10^{-3}$ & 0 \\
		$a_8 \, (\text{MeV fm}^8)$ & $-8.614 \times 10^{-6}$ & $-9.672 \times 10^{-6}$ & 0 \\
		$a_{10} \, (\text{MeV fm}^{10})$ & $1.621 \times 10^{-8}$ & $1.827 \times 10^{-8}$ & 0\\
		$b_0 \, (\text{MeV})$ & $34.70$ & $34.70$ & $32.43$ \\
		$b_2 \, (\text{MeV fm}^2)$ & -3.323 & -3.323 & 0 \\
		$b_4 \, (\text{MeV fm}^4)$ & $2.733\times 10^{-2}$ & $2.733\times 10^{-2}$ & 0 \\
		$b_6 \, (\text{MeV fm}^6)$ & $-8.992\ \times 10^{-5}$ & $-8.992\ \times 10^{-5}$ & 0 \\
		$b_8 \, (\text{MeV fm}^8)$ & $1.5849 \times 10^{-7}$ & $1.5849 \times 10^{-7}$ & 0 \\
		$b_{10} \, (\text{MeV fm}^{10})$ & $-1.738 \times 10^{-10}$ & $-1.738 \times 10^{-10}$ & 0\\ 	
		\bottomrule
	\end{tabular}
\end{table}

\subsection{Extended Skyrme effective interactions}
\label{Skyrme}
In this work, the baryon mean-field potentials in the LBUU model are derived from the N$5$LO Skyrme pseudopotential recently developed and implemented in Ref.~\cite{Wang:2024xzq}. Within the framework of N$5$LO Skyrme pseudopotential, the single-nucleon potential in symmetric nuclear matter and the (first-order) nuclear symmetry potential at saturation density $\rho_{0}$, i.e., $U_0$ and $U_{\rm sym}$, take the polynomial momentum-dependent forms
\begin{equation}
	\begin{aligned}
		\label{eq:U0_a}
		U_{0}(\rho_0,p) \equiv& a_{0}
		+ a_{2} \left(\frac{p}{\hbar}\right)^{2}
		+ a_{4} \left(\frac{p}{\hbar}\right)^{4} \\
		&+ a_{6} \left(\frac{p}{\hbar}\right)^{6}
		+ a_{8} \left(\frac{p}{\hbar}\right)^{8}
		+ a_{10} \left(\frac{p}{\hbar}\right)^{10},
	\end{aligned}
\end{equation}
and
\begin{equation}
	\begin{aligned}
		\label{eq:Usym_b}
		U_{\mathrm{sym}}(\rho_0,p) \equiv & b_{0}
		+ b_{2} \left(\frac{p}{\hbar}\right)^{2}
		+ b_{4} \left(\frac{p}{\hbar}\right)^{4} \\
		&+ b_{6} \left(\frac{p}{\hbar}\right)^{6}
		+ b_{8} \left(\frac{p}{\hbar}\right)^{8}
		+ b_{10} \left(\frac{p}{\hbar}\right)^{10},
	\end{aligned}
\end{equation}
where the coefficients $a_{n}$ and $b_{n}$ ($n=0,2,4,6,8,10$) are determined by the Skyrme parameters. This flexible parameterization allows one to explore systematically the influence of momentum dependence in the baryon mean field on the dynamics of HICs. In particular, the LBUU calculations based on the N5LO Skyrme pseudopotential~\cite{Wang:2024xzq} with empirical parameter values can reasonably reproduce proton flow data in Au+Au collisions at $\sqrt{s_{\rm{NN}}}=2.4~\rm{GeV}$ measured by HADES collaboration~\cite{HADES:2020lob,HADES:2022osk}, demonstrating the essential role of momentum dependence in the dynamics of HICs at GeV energies.

In the following LBUU simulations, we adopt the N5LO Skyrme pseudopotential SP$10$D$02$ reported in Ref.~\cite{Wang:2024xzq}
and denoted it by ``SP$10$'' here. In addition, based on the SP$10$, two new interactions, SP$10$-FL and SP$10$-MID, with different momentum dependence of single-nucleon potential are constructed following the strategy in Ref.~\cite{Wang:2024xzq}.
The $22$ adjustable parameters in the adopted extended Skyrme interaction are determined by $12$ characteristic quantities for the EOS of symmetric nuclear matter~(SNM) $E_0(\rho)$ and the symmetry energy $E_{\rm{sym}}(\rho)$, along with 10 coefficients $a_n$ and $b_n$ ($n=2,4,6,8,10$) in Eqs.~(\ref{eq:U0_a}) and~(\ref{eq:Usym_b}). The $12$ EOS characteristic quantities are defined by Taylor expansion coefficients of the $E_0(\rho)$ and $E_{\rm{sym}}(\rho)$ at saturation density $\rho_0$, including $\rho_0$,  $E_0(\rho_0)$, $K_0 =9\rho_0^2 [d^2E_0/d\rho^2]\vert_{\rho=\rho_0}$,
$J_{0}=27\rho_0^3 [d^3 E_0(\rho)/{d \rho^3}]|_{\rho=\rho_0}$,
$I_{0}=81\rho_0^4 [d^4 E_0(\rho)/{d \rho^4}]|_{\rho=\rho_0}$,
$H_0 =243\rho_0^5 [d^5 E_0(\rho)/{d \rho^5}]|_{\rho=\rho_0}$,
$E_{\rm{sym}}(\rho_0)$,
$L = 3\rho_0[dE_{\rm{sym}}(\rho)/d\rho]\vert_{\rho=\rho_0}$,
$K_{\rm{sym}} = 9\rho_0^2[d^2E_{\rm{sym}}(\rho)/d\rho^2]\vert_{\rho=\rho_0}$,
$J_{\rm{sym}} = 27\rho_0^3[d^3E_{\rm{sym}}(\rho)/d\rho^3]\vert_{\rho=\rho_0}$,
$I_{\rm{sym}} = 81\rho_0^4[d^4E_{\rm{sym}}(\rho)/d\rho^4]\vert_{\rho=\rho_0}$,
and $H_{\rm{sym}} = 243\rho_0^5[d^5E_{\rm{sym}}(\rho)/d\rho^5]\vert_{\rho=\rho_0}$.
For all the three interactions, we set $\rho_{0}$ = $0.160$ $\text{fm}^{-3}$, $E_{0}(\rho_0)$ = $-16.0$ \text{MeV}, $K_{0}$ = $230$ \text{MeV}, $J_{0}$ = $-383$ \text{MeV}, $I_{0}$ = $1818.9$ \text{MeV}, $H_{0}$ = $-12065$ \text{MeV}, $E_{\rm sym}(\rho_0)$ = $30$ \text{MeV}, $L$ = $45$ \text{MeV}, $K_{\rm sym}$ = $-110$ \text{MeV}, $J_{\rm sym}$ = $700$ \text{MeV}, $I_{\rm sym}$ = $-2458.5$ \text{MeV}, $H_{\rm sym}$ = $19663$ \text{MeV}, following SP$10$D$02$~\cite{Wang:2024xzq}. As for the coefficients $a_n$, they are tuned to fit the empirical optical potential from Hama et al. and its extrapolation up to kinetic energy of about $1.5$ GeV~\cite{Hama:1990vr,Cooper:1993nx} for SP$10$, while to reproduce the empirical optical potential by Feldmeier and Lindner~(FL)~\cite{Feldmeier:1991ey} for SP$10$-FL. Note the coefficients $b_n$ for both SP$10$ and SP$10$-FL are set to be the same as those for SP$10$D$02$~\cite{Wang:2024xzq}.

For SP$10$-MID, we set $a_{2}$-$a_{10}$ and $b_{2}$-$b_{10}$ to be zero, and determine $a_0$ and $b_0$ according to the HVH theorem~\cite{Hugenholtz:1958zz,SATPATHY199985,Chen:2011ag}, which results in momentum-independent single-nucleon potential in nuclear medium. Detailed values for $a_n$ and $b_n$ in the three extended Skyrme interactions are listed in Tab.~\ref{tab:sp_properties}.

\begin{figure}[t]
	\centering
	\includegraphics[width=\linewidth]{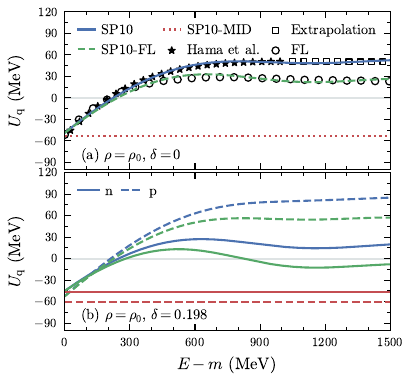}
	\caption{Energy dependence of single-nucleon potential $U_q$ in cold nuclear matter predicted by SP$10$, SP$10$-FL and SP$10$-MID. (a) Results for symmetric nuclear matter ($\delta=0$) at saturation density $\rho_0$  are compared with the nucleon optical potential in  by Hama {\it et al.}\cite{Hama:1990vr} (solid stars) and its extrapolation to $1.5$ GeV (open squares), as well as with the result from Feldmeier and Linder (FL)~\cite{Feldmeier:1991ey} (open circles). (b) Neutron and proton potentials  in asymmetric nuclear matter at density $\rho=\rho_0$ and  isospin asymmetry $\delta=0.198$, a value typical for a $^{197}\text{Au}$ nucleus.}
	\label{fig:U0}
\end{figure}

Figure~\ref{fig:U0} (a) compares the energy dependence of single-nucleon potential in SNM at $\rho_0$ predicted by the three interactions with nucleon optical potential extracted from proton-nucleus scattering data~\cite{Hama:1990vr}. It is seen that the SP$10$ and SP$10$-FL well reproduce the optical potential by Hama {\it et al.}\cite{Hama:1990vr,Cooper:1993nx} as well as Feldmeier and Linder~\cite{Feldmeier:1991ey}, respectively. The empirical optical potential from Hama {\it et al.} and its extrapolation up to kinetic energy of about $1.5$ GeV~\cite{Hama:1990vr,Cooper:1993nx} corresponds to the Schrödinger-equivalent potential based on the Dirac phenomenology. The optical potential from Feldmeier and Lindner (FL) in Ref.~\cite{Feldmeier:1991ey} employs a different definition of the non-relativistic optical potential optical based on the same Dirac phenomenology of proton scattering data by Hama {\it et al.}~\cite{Hama:1990vr}. The different definitions for non-relativistic optical potential lead to distinct energy dependence. We can see that the Hama potential predicts larger optical potentials at high energies, which results in stronger momentum dependence in SP$10$ than that in SP$10$-FL. For SP$10$-MID, the single nucleon potential is momentum/energy independent. As a result, it takes a constant value of $U_0 = -52.84$ MeV in SNM at $\rho_0$.

Figure~\ref{fig:U0} (b) shows proton and neutron potentials  in asymmetric nuclear matter at density  $\rho =\rho_0$ and isospin  asymmetry of $\delta=0.198$, comparable to that of a $^{197}\text{Au}$ nucleus. As expected, the SP$10$-MID interaction, being momentum independent, yields constant neutron and proton potentials. In contrast, both SP$10$ and SP$10$-FL predict a noticeably stronger momentum dependence for protons than for neutrons, reflecting the positive neutron-proton effective-mass splitting in neutron-rich matter.

For the mean-field potentials of $\Delta$ resonances,
we note that the isospin dependence of the $\Delta$ single-particle potential remains an open question~\cite{Cai:2015hya,Cozma:2014yna,Drago:2014oja,Li:2015hfa}. There are two popular forms: one proposed in Ref.~\cite{Li:2002yda} and the other in Ref.~\cite{UmaMaheswari:1997mc}, and both are expressed as a linear scaling of neutrons ($n$) and protons ($p$) potentials. In this work, following Ref.~\cite{Wang:2024xzq}, we adopt the relations between $\Delta$ and nucleon potentials proposed in Ref.~\cite{UmaMaheswari:1997mc}, namely,
$U_{\Delta^{++}} = -U_n + 2U_p, U_{\Delta^{+}} = U_p, U_{\Delta^{0}} = U_n, U_{\Delta^{-}} = 2U_n - U_p$. As for higher resonances and pions, they are treated as freely propagating particles without mean-field potentials.

\begin{figure}[t]
	\centering
	\includegraphics[width=0.95\linewidth]{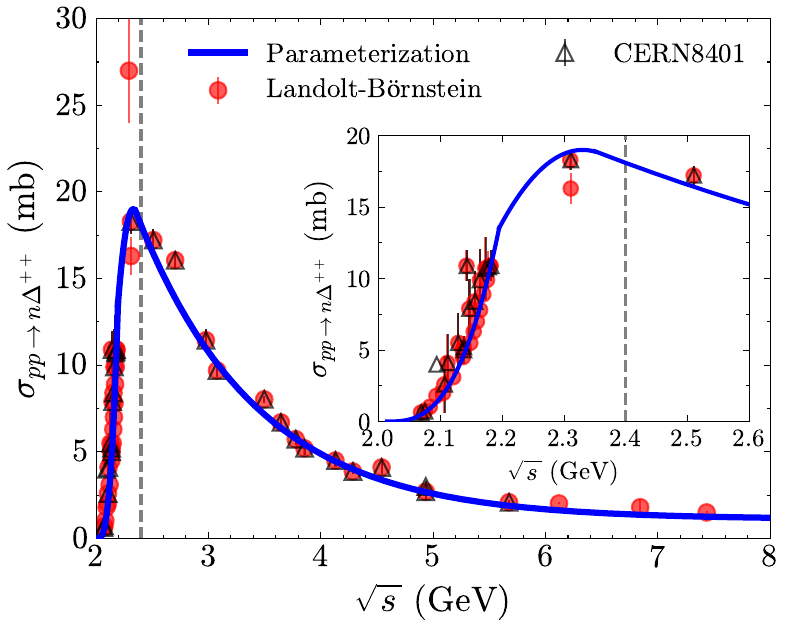}
	\caption{Cross section for $pp \rightarrow n\Delta^{++}$ as a function of the invariant mass $\sqrt{s}$ (in GeV). Experimental data are taken from CERN8401~\cite{Flaminio:1984gr} (black hollow triangles) and Landolt-Börnstein~\cite{Schopper:1988hwx} (Red solid dot). The solid blue line represents the parameterization given by Eq.~(\ref{eq:NNND}). The inset provides an enlarged view of the cross section near threshold, and the black dashed line indicates  $\sqrt{s}$ = 2.4 GeV.}
	\label{fig:NNNDpara}
\end{figure}

\subsection{Modeling of pion production}
\label{2-3NNND}
In the present LBUU model, pions are produced through the decays of  $\Delta$, $N^*$ and $\Delta^*$ resonances.
Since $\Delta$ decay dominates the final $\pi$ yield in HICs at intermediate energies, an accurate description of the $N+N \rightarrow N+\Delta$ cross section is critical for transport-model studies of pion production. In this work, we introduce a new parameterization for $p+p\rightarrow n+\Delta^{++}$ cross sections by fitting  experimental data from Refs.~\cite{Flaminio:1984gr} ( CERN8401) and~\cite{Schopper:1988hwx} (Landolt-Börnstein), i.e.,
\begin{equation}
	\label{eq:NNND}
	\sigma_{pp \rightarrow n\Delta^{++}}(\sqrt{s}) =
	\begin{cases}
		1650 \cdot (\sqrt{s} - 2.014)^{2.81}, \\
		\quad \text{for} \ 2.014 \ \text{GeV} < \sqrt{s} \leq 2.20 \ \text{GeV} \\
		19 - 300 \cdot (\sqrt{s} - 2.33)^2, \\
		\quad \text{for} \ 2.20 \ \text{GeV} \leq \sqrt{s} \leq 2.35 \ \text{GeV} \\
		1.104 + 17.682 \cdot \exp\left(-\frac{\sqrt{s} - 2.357}{0.08}\right), \\
		\quad \text{for} \ \sqrt{s} \geq 2.35 \ \text{GeV}
	\end{cases}
\end{equation}
where the invariant mass $\sqrt{s}$ and the cross section $\sigma_{pp\rightarrow n \Delta^{++}}$
are in units of GeV and mb, respectively. As shown in Fig.~\ref{fig:NNNDpara}, this parametrization well reproduces the experimental data from threshold up to high energies of $\sim 7~\rm{GeV}$. Cross sections for other $\Delta$ production channels are related by isospin symmetry as
$\sigma_{nn\rightarrow p+\Delta^{-}}(\sqrt{s}) =\sigma_{pp \rightarrow n\Delta^{++}}(\sqrt{s})$,   $\sigma_{pp\rightarrow p+\Delta^{+}}=
\sigma_{nn \rightarrow n+\Delta^{0}} = \frac{1}{3}\sigma_{pp \rightarrow n\Delta^{++}}(\sqrt{s})$, and $\sigma_{np\rightarrow n+\Delta^{+}}=
\sigma_{np\rightarrow p+\Delta^{0}} = \frac{2}{3}\sigma_{pp \rightarrow n\Delta^{++}}(\sqrt{s})$. The $\Delta$ absorption ($\sigma_{N\Delta\rightarrow NN}$ ) cross sections are obtained from the corresponding production cross sections by detailed balance condition~\cite{Danielewicz:1991dh}.
For the $\Delta$ resonance, we adopt the spectral function, decay width, and absorption cross section from Ref.~\cite{TMEP:2023ifw}. The corresponding properties for the $\Delta^*$ and $N^*$ resonances are taken from Ref.~\cite{SMASH:2016zqf}.

It should be noted that, in principle, momentum-dependent mean-field potentials affect not only the reaction dynamics but also the kinematics and the in-medium cross sections and decay widths entering all inelastic scattering and decay channels. In the present work, however, we focus solely on the dynamical impact of the momentum dependence of the nucleon potentials and neglect mean-field potentials in treating collisions. The impact of this momentum dependence on the collision terms, such as threshold effects~\cite{Ferini:2005del,Song:2015hua} or in-medium corrections to the $NN$ inelastic cross sections~\cite{Godbey:2021tbt,Kummer:2023hvl}, is left for future studies.

To evaluate the momentum dependence effects on pion production, we solve the BUU equation with the three interactions, i.e., SP$10$, SP$10$-FL and SP$10$-MID to simulate the Au+Au collisions at $\sqrt{s_{\rm{NN}}}=2.4$ GeV. Some details of the simulations are as follows: the lattice spacing is set to be $1$ fm; the gradient parameter $E^{[2]}$ in the Thomas-Fermi initialization for the ground state of $^{197}$Au as described in Refs.~\cite{Wang:2024xzq,Wang:2019ghr} is set to $-310$ MeV fm$^{5}$ for the interactions SP$10$, SP$10$-FL and SP$10$-MID, to fit the experimental binding energy of $^{197}$Au; the number of test particles is $50,000$; the simulation of the reaction is stopped at $60$ fm/$c$, with a time step of $0.2$ fm/$c$; the free nucleon-nucleon elastic cross section $\sigma_{\rm{NN}}^{\text{free}}$
is based on the parametrization of experimental nucleon-nucleon scattering data~\cite{Cugnon:1996kh}, and an in-medium correction to $\sigma_{\rm{NN}}^{*}$ is parameterized according to the nuclear giant dipole resonance width~\cite{Wang:2020xgk}.

\begin{figure}[t]
	\centering
	\includegraphics[width=0.95\linewidth]{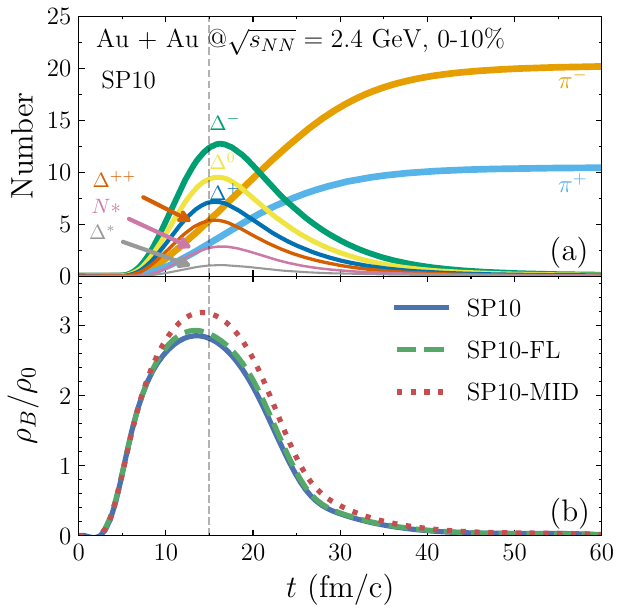}
	\caption{Time evolution of (a) the number of pions ($\pi^{-}, \pi^{+}$) and resonances ($\Delta^{-}$, $\Delta^{0}$, $\Delta^{+}$, $\Delta^{++}$, $N^{*}$, $\Delta^*$ )  obtained from LBUU calculation using the SP$10$ interaction, and (b) the central baryon density $\rho_{B}$  predicted by the SP$10$, SP$10$-FL and SP$10$-MID. Both panels correspond to the Au+Au collisions at $\sqrt{s_{\rm{NN}}} = 2.4 \, \text{GeV}$
		and  $0$-$10\%$ centrality. The vertical black dashed lines indicate $t$ = 15 fm$/c$.}
	\label{fig:timeyldrhocenter}
\end{figure}

\section{Results and discussion}
\label{Sec3}
Figure~\ref{fig:timeyldrhocenter}~(a) shows the time evolution of the number of charged pions, $\Delta$-resonances, and higher-lying resonances in
Au+Au collisions at $\sqrt{s_{\rm{NN}}}=2.4~\rm{GeV}$ and centrality
$0-10\%$ (impact parameter $b=0-4.7~\rm{fm}$) calculated using the SP$10$ interaction. As expected,  the $\Delta$ resonances are produced in much more abundant than the higher-lying $N^*$ and $\Delta^*$, and therefore dominate the production of charged pions.
Fig.~\ref{fig:timeyldrhocenter}~(b) displays the time evolution of the baryon density $\rho_B$ in a central cell with $1\times1\times1~\rm{fm}^3$ obtained from the LBUU simulations using SP$10$, SP$10$-FL and SP$10$-MID. By comparing panels (a) and (b) of Fig.~\ref{fig:timeyldrhocenter}, one clearly sees that these resonances and pions are produced during the compressing stage. Both the central baryon density and resonance numbers reach their maxima around $t=15~\rm{fm}/c$. For the three interactions, SP$10$, SP$10$-FL and SP$10$-MID, the maximum baryon densities are about $2.9\rho_{0}$, $2.95\rho_{0}$ and $3.2\rho_0$, respectively. We can see that stronger momentum dependence, as in SP$10$, not only reduces the maximum achievable density but also causes it to decrease more rapidly over time. This suggests that a stronger momentum dependence is associated with a more anisotropic compression with a larger squeeze-out pressure (and thus a larger magnitude of nucleon elliptic flows), which enhances particle emission and shortens the compression stage.

\begin{table}[bt]
	\caption{Predicted charged pion multiplicities for the $0$-$10\%$ and $20$-$30\%$ centrality bins from the LBUU model using the SP$10$, SP$10$-FL and SP$10$-MID interactions. Experimental data measured by HADES collaboration are from Ref.~\cite{HADES:2020ver}.\label{Tab:pion}}
	\centering
	\begin{tabular}{lcccc}
		\toprule
		\multirow{2}{*}{} & \multicolumn{2}{c}{$0$--$10\%$} & \multicolumn{2}{c}{$20$--$30\%$} \\
		\cmidrule(lr){2-3} \cmidrule(lr){4-5}
		& $\pi^-$ & $\pi^+$ & $\pi^-$ & $\pi^+$ \\
		\midrule
		HADES &$17.2(11)$ & $9.3(7)$ &$8.7(6)$ & $4.7(3)$  \\
		SP10 &20.3 &10.5 &9.6 &4.7\\
		SP10-FL &21.3 & 11.1 &10.2 & 5.0\\
		SP10-MID &25.6 &14.3&12.5 & 6.7\\
		\bottomrule
	\end{tabular}
\end{table}

\begin{figure}[bt]
	\centering
	\includegraphics[width=0.95\linewidth]{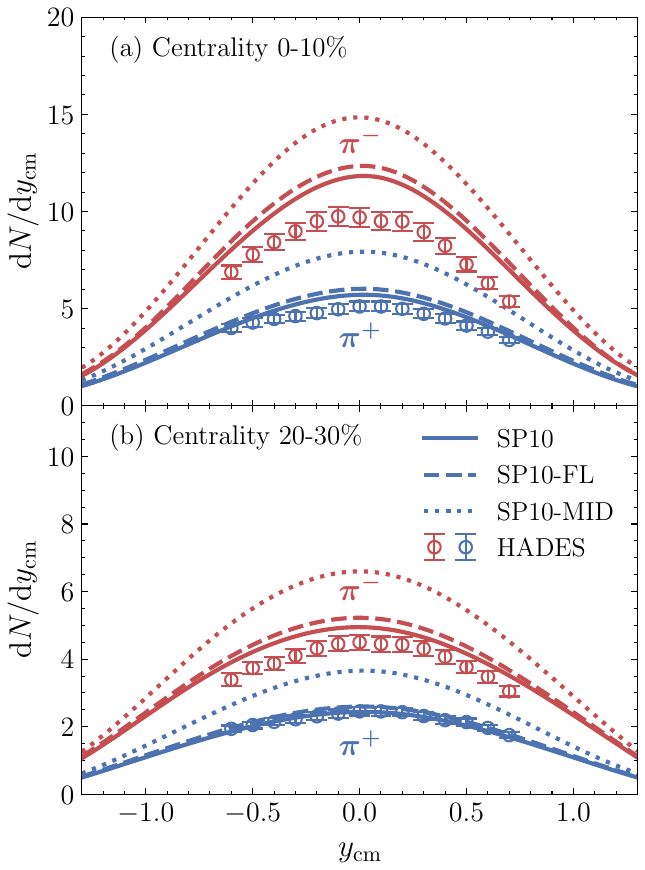}
	\caption{Rapidity distributions of pions with different momentum dependence interactions: SP$10$, SP$10$-FL, and SP$10$-MID in Au + Au collisions at $\sqrt{s_{\rm{NN}}} = 2.4 \, \text{GeV}$. Panel (a) shows the results for centrality $0-10$$\%$ and panel (b) for centrality 20-30$\%$. The solid, dashed and dotted curves represent LBUU model predictions, and the symbols represent the HADES experimental data~\cite{HADES:2020ver} with error bars indicating uncertainties.}.
	\label{fig:HADESdndy}
\end{figure}

\begin{figure}[ht]
	\centering
	\includegraphics[width=0.95\linewidth]{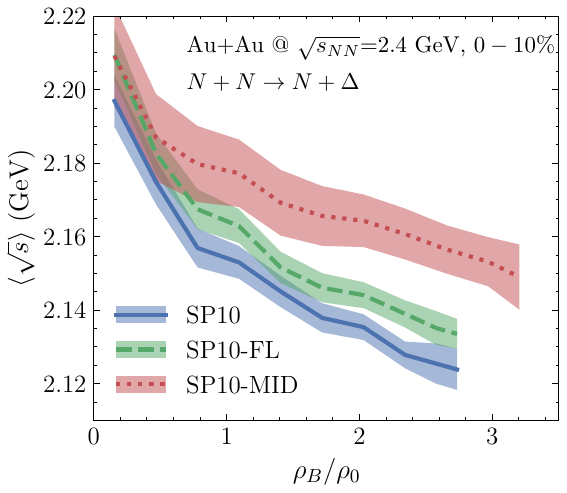}
	\caption{The averaging invariant mass $\langle\sqrt{s} \rangle$ in $N+N\rightarrow N+\Delta$ reaction as functions of baryon density $\rho_{B}$ at the evolution time of $t=15~\rm{fm}/c$ obtained from the LBUU simulations
		with the SP$10$, SP$10$-FL, and SP$10$-MID interactions, respectively. Statistical errors are displayed as the shaded bands.}
	\label{fig:sqrts}
\end{figure}

In Figs.~\ref{fig:HADESdndy}~(a) and~(b), we further show the rapidity distributions of charged pions in the $0$-$10\%$ and $20-30$$\%$ centrality bins predicted by the LBUU simulations with the SP$10$, SP$10$-FL and SP$10$-MID interactions. For reference, HADES data~\cite{HADES:2020ver} are plotted as open circles, and the corresponding integrated yields are listed in Tab.~\ref{Tab:pion}.  A comparison of results from the three interactions shows the pion yield is strongly sensitive to the momentum-dependence of the nucleon mean-field potential. In particular, in central (centrality $0$–$10 \%$) collisions, the momentum-independent SP$10$-MID interaction significantly enhances pion production, giving yields about $29.6 \%$ and $23.1 \%$ larger than those of SP$10$ and SP$10$-FL, respectively. While the difference between SP$10$ and SP$10$-FL remains modest, about $5.8 \%$. These results indicate that stronger momentum dependence systematically suppresses pion production. The overall consistency of the SP$10$ predictions with the HADES measurements seems to further supports the need for a stronger momentum-dependent nucleon potential.

The sensitivity of pion yields to the momentum dependence of the nucleon mean-field potential can be understood from the impact of the momentum dependence on nucleon dynamics. As shown in Fig.~\ref{fig:timeyldrhocenter}, a stronger momentum dependence lowers both the maximum density of the compressed nuclear matter formed in the collision and the duration of the compression stage, thereby reducing the number of binary collisions and consequently pion production. More importantly, the momentum-dependent potential decreases the kinetic energy of a nucleon as the nucleon moves into dense nuclear medium, leading to a smaller fraction of high-momentum nucleons during compression. To illustrate this mechanism, we analyze the single time step centered at $t=15~\rm{fm}/c$ and calculate
the average invariant mass $\langle \sqrt{s} \rangle$ for $N+N\rightarrow N+\Delta$ reaction as a function of local baryon density $\rho_{B}$ in the LBUU simulations  with the three interactions, SP$10$, SP$10$-FL and SP$10$-MID. For each of the colliding nucleon pairs in that time step, we compute $s = (p_{1,\rm{free}}+p_{2,\rm{free}})^2$ with free momenta $p_{i,\rm{free}} =(\sqrt{m^2+\bm{p}_i^2}, \bm{p}_i)$, and record the local $\rho_B$ at the collision point. The pairs are then grouped by $\rho_B$, and $\langle\sqrt{s}\rangle$ is evaluated within each density bin.
The results are shown in Fig.~\ref{fig:sqrts} for central Au+Au collisions (centrality $0$-$10 \%$). One can see that stronger momentum dependence results in a lower average invariant mass in nucleon-nucleon~($NN$) scattering. Since the $NN$ inelastic scattering cross section is highly sensitive to the invariant mass near threshold (see Fig.~\ref{fig:NNNDpara}), this effect further suppresses pion production.

As we have mentioned before, the pion production in intermediate-energy HICs involves a variety of complex mechanisms. A fully quantitative description of pion production and a reliable extraction of nucleon dynamics from pion observables require more sophisticated and systematic modeling. In fact, one sees from Fig.~\ref{fig:HADESdndy} that the SP$10$ still over-predicts the $\pi^{-}$ yield, especially for the central collisions of $0$-$10\%$ centrality. Nevertheless, we would like to emphasize that the present work does not attempt a comprehensive quantitative study; instead, we focus on a qualitative investigation of the influence of the momentum dependence of nucleon mean-field potentials on the pion yield.

Finally, we note that although the mechanism discussed here differs from that of isospin fractionation in Refs.~\cite{Li:2000bj,Li:2002yda}, there is a conceptual analogy. Isospin fractionation is driven by the density dependence of the symmetry energy and leads to a spatial separation of neutrons and protons. In contrast, the effect emphasized in this work originates from the momentum dependence of the nucleon mean-field potential, which modifies the fraction of high-momentum nucleons and thus alters the reaction dynamics in heavy-ion collisions. This change in the fraction of high-momentum nucleons can leave clear signatures in various observables, most notably in the final pion yields. For subthreshold particle production, the effect is expected to be particularly important when the nucleon mean-field potential exhibits a strong momentum dependence.

\section{Summary and outlook}
\label{Sec4}
We have investigated how the momentum dependence of the nucleon mean-field potential influences pion production in intermediate-energy HICs by using the lattice Boltzmann-Uehling-Uhlenbeck transport model with extended Skyrme interactions named N5LO Skyrme pseudopotential. Three representative interactions, SP$10$, SP$10$-FL, and SP$10$-MID, are constructed to span the range from strong to vanishing momentum dependence.
Employing an updated $NN\rightarrow N\Delta$ cross section and simulating Au+Au reactions at $\sqrt{s_{\rm{NN}}}=2.4$ GeV, we find that stronger momentum dependence systematically suppresses pion yields, while weaker or momentum-independent mean fields enhance them.

A detailed dynamical analysis reveals that the suppression of the pion yield originates from the modification of nucleon momentum distributions, namely, a stronger momentum dependence in the nucleon mean-field potentials lowers the fraction of high-momentum nucleons during the compression in heavy-ion collisions, thereby reducing the average invariant mass in nucleon-nucleon inelastic scattering and, consequently, the production of $\Delta$-resonances and pions.

These findings underscore the importance of the momentum-dependence of nucleon potentials in the studies of intermediate-energy HICs. They further highlight mean-field-induced modification of nucleon high-momentum phase space as a key microscopic mechanism that links the momentum dependence of the nuclear mean field to pion observables and, ultimately, to constraints on the high-density nuclear matter equation of state.

\section*{Acknowledgements}
This work was supported in part by the National Natural Science Foundation of China under Grant Nos. 12235010 and 12147101, the National SKA Program of China (Grant No. 2020SKA0120300), the Science and Technology Commission of Shanghai Municipality (Grant No. 23JC1402700), and the Natural Science Foundation of Henan Province (Grant No. 242300421048).



\bibliographystyle{elsarticle-num}
\bibliography{HADESpion}






\end{document}